\documentclass[journal]{IEEEtran}
\usepackage{graphicx}

\ifCLASSINFOpdf
\else
\fi
\begin{document}
%
\title{Optical and Radiographical Characterization of Silica Aerogel for Cherenkov Radiator}
%
%
%

\author{Makoto Tabata,
              Ichiro Adachi,
              Yoshikiyo Hatakeyama,
              Hideyuki Kawai,
              Takeshi Morita,
              and Keiko Nishikawa
\thanks{M. Tabata is with the Institute of Space and Astronautical Science (ISAS), Japan Aerospace Exploration Agency (JAXA), Sagamihara 252-5210, Japan and with the Department of Physics, Chiba University, Chiba 263-8522, Japan (corresponding author to provide e-mail: makoto@hepburn.s.chiba-u.ac.jp).}
\thanks{I. Adachi is with the Institute of Particle and Nuclear Studies (IPNS), High Energy Accelerator Research Organization (KEK), Tsukuba 305-0801, Japan.}
\thanks{Y. Hatakeyama, T. Morita, and K. Nishikawa are with the Graduate School of Advanced Integration Science, Chiba University, Chiba 263-8522, Japan.}
\thanks{H. Kawai is with the Department of Physics, Chiba University, Chiba 263-8522, Japan.}
\thanks{Manuscript received February 29, 2012; revised May 15, 2012. This work was supported in part by the Japan Society for the Promotion of Science (JSPS) under Grants No. 07J02691 (for M.T.) and 19540317 (for I.A.).}}

%
%

\markboth{Preprint submitted to IEEE TRANSACTIONS ON NUCLEAR SCIENCE}%
{Shell \MakeLowercase{\textit{et al.}}: Bare Demo of IEEEtran.cls for Journals}
%



\maketitle

\begin{abstract}
We present optical and X-ray radiographical characterization of silica aerogels with refractive index from 1.05 to 1.07 for a Cherenkov radiator. A novel pin-drying method enables us to produce highly transparent hydrophobic aerogels with high refractive index by shrinking wet-gels. In order to investigate the uniformity in the density (i.e., refractive index) of an individual aerogel monolith, we use the laser Fraunhofer method, an X-ray absorption technique, and Cherenkov imaging by a ring imaging Cherenkov detector in a beam test. We observed an increase in density at the edge of the aerogel tiles, produced by pin-drying.
\end{abstract}


%
\IEEEpeerreviewmaketitle

\section{Introduction}
%
%
%
%
\IEEEPARstart{S}{ilica} aerogel is one of the most important Cherenkov radiators owing to its unique refractive index. Aerogels have been used as a radiator for threshold Cherenkov counters in many high-energy and nuclear experiments (e.g., \cite{cite1}), and so far, conventionally produced aerogel tiles have been regarded as having sufficiently uniform internal refractive index for this purpose. However, it is crucial to investigate the uniformity of refractive index over an individual aerogel monolith. In particular, when we use aerogels as a radiator for ring imaging Cherenkov (RICH) counters, the uniformity significantly affects the detector performance because the Cherenkov angle of emitted photons is determined by the local refractive index of the aerogels \cite{cite2}--\cite{cite4}. Aerogel-based RICH counters have already been used in several experiments \cite{cite5}--\cite{cite7}. To develop a proximity focusing RICH counter using a dual-refractive-index aerogel radiator \cite{cite8} for the Belle II experiment \cite{cite9}, we have been attempting to produce highly transparent hydrophobic aerogels with high refractive index ($n >$ 1.04). We recently discovered a new technique, the pin-drying method, for producing aerogels with excellent transparency \cite{cite10}--\cite{cite12}. The method is designed such that a synthesized wet-gel shrinks in the pin-drying process to increase its transparency before supercritical drying. Because of the effect on performance of various types of Cherenkov detectors, the uniformity of refractive index within the shrunken aerogel monolith must be examined in detail.

To investigate the internal tile uniformity of aerogels produced by both pin-drying and conventional methods, we used three methods: the laser Fraunhofer method, an X-ray absorption technique, and Cherenkov imaging using a RICH counter in a beam test. The Fraunhofer method is the most readily available method of measuring the refractive index directly but it can be applied only at the corner of the aerogel tiles. Well-focused X-rays are the most promising probe; using them, we can scan aerogel tiles point by point to obtain the density distribution utilizing their absorption \cite{cite13}. Measuring the density is equivalent to evaluating the refractive index because the refractive index of aerogels is proportional to their density $\rho $:
\begin{equation}
\label{eq:eq1}
n = 1+k\rho ,
\end{equation}
where $k$ is a constant \cite{cite14}. Because X-ray absorption is very sensitive to aerogel thickness, we had to cut the aerogels to evaluate the thickness distribution along the line scanned by X-rays. By using charged beam with high momentum and a RICH counter, which is composed of position-sensitive photodetectors, we also measured the Cherenkov angle of photons emitted along each beam track on an aerogel tile. The radiator performance in RICH counters can be evaluated more directly by this method; however, the opportunity for beam test experiments is limited. Consequently, only two aerogel tiles were measured in this study. The purpose of this paper is the optical and X-ray radiographical characterization of aerogels with a relatively small tile size from trial production. We discussed the refractive index uniformity in the planar direction without considering the thickness direction of the aerogels. This is because we aim at developing the large area (3.5 m$^2$) RICH counter for Belle II. One of the demands for the Cherenkov radiator is sufficient uniformity over the whole area of the counter to obtain the constant mean of the Cherenkov angle distribution depending on the velocity of charged particles. The accumulated effect of the refractive index uniformity in the thickness direction would be included in the investigation of the planar direction.

In this paper, the methods for producing silica aerogel and its transparency characteristics are briefly summarized in Section II. Sections from III to V describe the three approaches for evaluating the aerogel tile uniformity and the measurement results in detail. Finally, we discuss the quality of our aerogels by comparing the results of different measurements in Section VI and conclude these characterization analyses of aerogel Cherenkov radiators in Section VII.

 

\section{Production Methods}

\renewcommand\thefootnote{\alph{footnote}}

\begin{table*}[!t]
\centering
\caption{List of Aerogels}
\label{table:table1}
\begin{minipage}{11.8cm}
\renewcommand{\arraystretch}{1.25}
	\begin{tabular}{cccccc}
		\hline
		ID (our reference) & $n_{tag}$ & Method & $\Lambda_T$\footnote{Transmission length.} (mm) & $\rho$ (g/cm$^3$) & Dimensions (mm$^3$) \\
		\hline
		C-1.045 (PDR21-3a) & 1.0448 & Conventional & 46.5 & 0.156 & 95 $\times $ 95 $\times $ 20.7 \\
		C-1.050 (PDR1-4b) & 1.0507 & Conventional & 48.1 & 0.177 & 92 $\times $ 92 $\times $ 20.9 \\
		P-1.055 (PDR20-1a) & 1.0538 & Pin-drying & 48.4 & 0.184 & 91 $\times $ 91 $\times $ 19.8 \\
		C-1.060 (PDR21-5a) & 1.0606 & Conventional & 30.5 & 0.212 & 94 $\times $ 94 $\times $ 20.6 \\
		P-1.060 (PDR11-2a) & 1.0604 & Pin-drying & 57.1 & 0.200 & 88 $\times $ 88 $\times $ 19.6 \\
		P-1.065 (PDR1-6a) & 1.0646 & Pin-drying & 52.8 & 0.218 & 86 $\times $ 86 $\times $ 19.8 \\
		P-1.070 (PDR18-1a) & 1.0717 & Pin-drying & 46.1 & 0.246 & 82 $\times $ 82 $\times $ 19.6 \\
		\hline
	\end{tabular}
\end{minipage}
\end{table*}

\begin{figure}[t]
\centering
\includegraphics[width=3.5in]{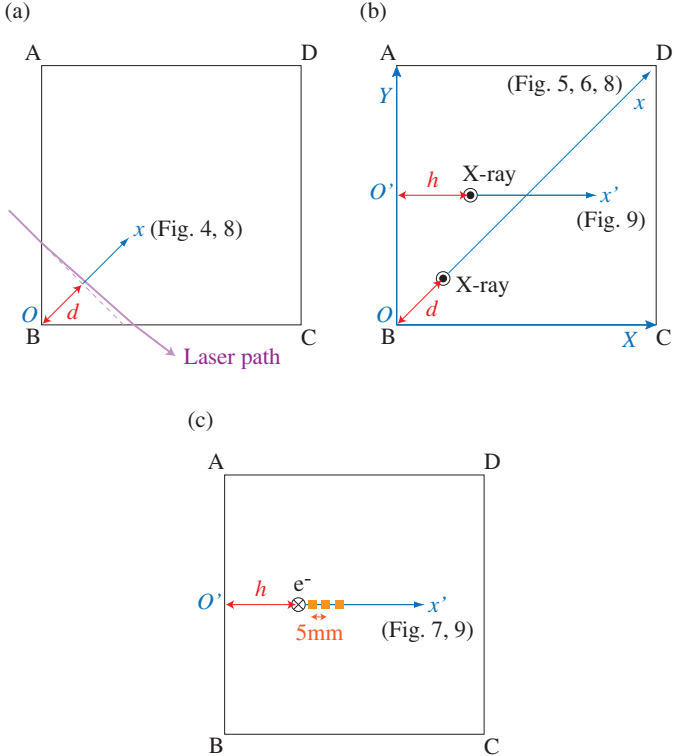}
\caption{Geometrical coordinates on the face of an aerogel tile. (a) For the Fraunhofer method (Section III). The corner $B$ is regarded as the origin $O$ and the diagonal $x$-axis is defined toward the tile center. The symbol $d$ shows the minimum distance from the vertex $B$ to the laser path (dashed line without the aerogel). (b) For X-ray absorption measurements (Section IV). X-ray scans were performed along the $x$ and $x'$-axis. For the diagonal scan, the $x$-axis and $d$ are defined in the same way as (a), namely $d$ is the distance from $O$ to the X-ray incident position. For the horizontal scan, the $x'$-axis is defined toward the tile center when the midpoint of the side $AB$ is regarded as the origin $O'$. The symbol $h$ shows the distance from the edge $AB$ to the X-ray. The $X$--$Y$ coordinate is employed when an aerogel tile is placed in a holder with a movable stage. (c) For a beam test (Section V). The $x'$-axis and $h$ are defined in the same way as (b), namely $h$ is the distance from $O'$ to the electron track position. Filled squares at 5-mm increments denote the track accumulation areas (3 mm $\times $ 3 mm) for analysis.}
\label{fig:fig1}
\end{figure}

\begin{figure}[t]
\centering
\includegraphics[width=3.5in]{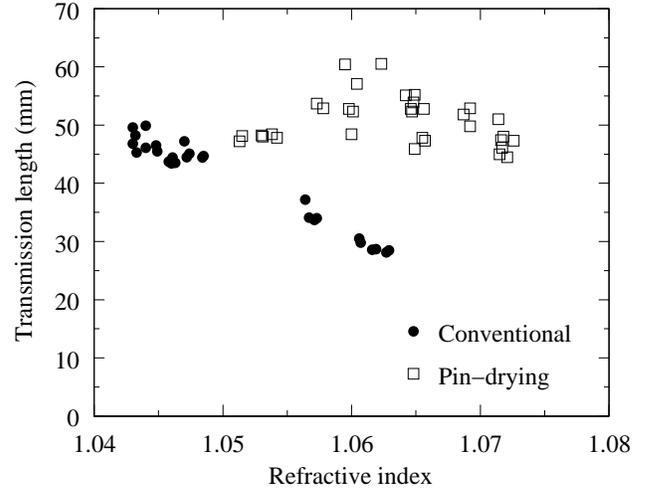}
\caption{Transmission length as a function of the refractive index for 60 aerogels produced by the conventional (circles) and pin-drying (squares) methods. The transmission length was evaluated at $\lambda$ = 400 nm, and the refractive index was measured at $\lambda$ = 405 nm.}
\label{fig:fig2}
\end{figure}

Sample tiles of silica aerogel were prepared by the conventional or pin-drying production methods with \textit{N,N}-dimethylformamide (DMF) as a solvent. The wet-gels with a thickness of approximately 20 mm were synthesized using a square mold of 96-mm side. The four corners of an aerogel tile were named as $A$, $B$, $C$, and $D$ in counterclockwise direction. Fig. \ref{fig:fig1} shows the geometrical coordinates on the aerogel tile used throughout this paper. The samples described in this paper are listed in Table \ref{table:table1}. The tagged refractive index ($n_{tag}$) is defined in Section III. The average density was determined gravimetrically.

In our conventional method (developed at KEK, Japan), a synthesized wet-gel was aged and then subjected to hydrophobic treatment. Finally, the wet-gel was converted to an aerogel by CO$_2$ supercritical drying. The detailed procedure for producing the aerogel is given in \cite{cite14}. A minor shrinkage (from 0.96 to 0.99) in length for each side was observed in the conventional production method.

In the pin-drying method, a wet-gel was prepared in the same way as for the production of conventional aerogels with $n$ = 1.049. The aged wet-gel was then enclosed in a special container with pinholes before hydrophobic treatment. The solvent vapor contained in the wet-gel gradually evaporated through the pinholes for several weeks, which shrank the wet-gel without cracking it. Thus, this method yields wet-gels with higher density. After pin-drying, hydrophobic treatment and CO$_2$ supercritical drying were performed. The shrinkage ratio in length ranged from 0.85 to 0.95 for the aerogels listed in Table \ref{table:table1} (pin-drying). The pin-drying method is described in greater detail in \cite{cite12}.

\begin{figure}[t]
\centering
\includegraphics[width=3.5in]{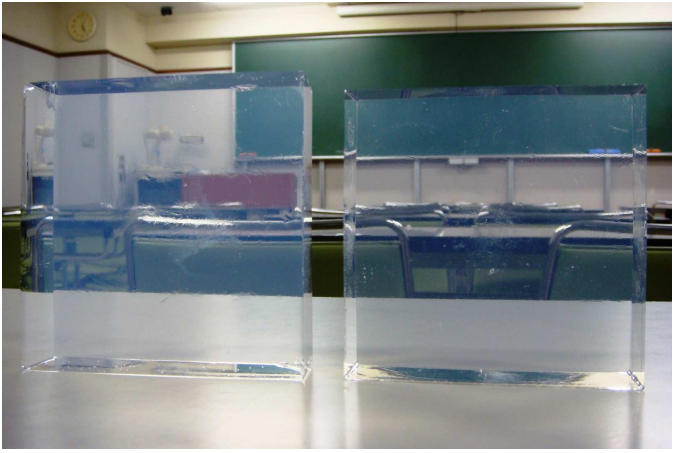}
\caption{Photograph of aerogels produced by the conventional (left, $n_{tag}$ = 1.063, $\Lambda_T$ = 28.5 mm, our reference: PDR21-7b) and pin-drying (right, $n_{tag}$ = 1.065, $\Lambda_T$ = 45.9 mm, our reference: BTR11-3a) methods.}
\label{fig:fig3}
\end{figure}

Fig. \ref{fig:fig2} shows the transmission length ($\Lambda_T$) as a function of the refractive index for the aerogels produced by the conventional and pin-drying methods. The transmission length was calculated using the aerogel thickness and the transmittance at 400 nm wavelength ($\lambda$). The refractive index was measured at $\lambda$ = 405 nm by the Fraunhofer method. Aerogels having a long transmission length can be produced in the range $n >$ 1.05 by pin-drying. At $n$ = 1.06, the transmission length of aerogels produced by pin-drying was approximately twice that of conventionally fabricated aerogels. A photograph of aerogels produced by the conventional and pin-drying methods is shown in Fig.\ref{fig:fig3}. The thickness of the samples is 2 cm.

\section{Laser Fraunhofer Method}

First, we evaluated the uniformity of the aerogels' refractive index by the Fraunhofer method with a blue--violet ($\lambda$ = 405 nm) semiconductor laser. In this method, the refractive index at the corner of an aerogel tile is obtained by measuring the minimum angle of deviation of the laser. Manufactured aerogels are usually tagged with the refractive index measured at $d$ = 5 mm, where $d$ is the minimum distance from each vertex of the aerogel to the laser's path without the aerogel (Fig. \ref{fig:fig1}a). Moreover, the standard method of refractive index measurement is described in \cite{cite14}. By changing $d$, the refractive index at each point distant from the vertex can be measured. However, the effect of accumulated refraction along the laser's path is included. The laser spot was 3.5 mm in diameter at the aerogel position but was focused on a screen to measure its displacement. The focused laser spot was less than 1 mm in diameter in the absence of the aerogel. The laser path lay in the mid-plain of the aerogel tile.

\begin{figure}[t]
\centering
\includegraphics[width=3.5in]{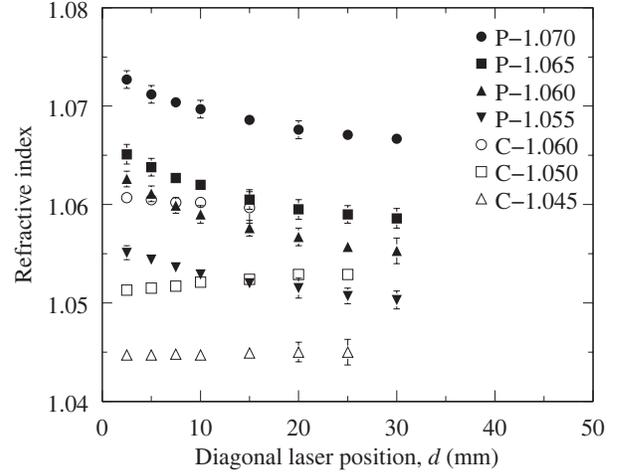}
\caption{Refractive index as a function of the diagonal laser position $d$ for aerogels of different refractive indices produced by pin-drying (filled symbols) or conventional (open symbols) methods. Aerogels are tagged with the mean refractive index and are represented by circles, squares, triangles, and inverted triangles. Owing to the low transparency of the conventionally produced aerogels, the plot lacks data at positions away from the vertex of the aerogels; i.e., the laser could not pass through the aerogels at these positions.}
\label{fig:fig4}
\end{figure}

Fig. \ref{fig:fig4} shows the refractive index as a function of the diagonal laser position $d$ for seven aerogels produced by the conventional or pin-drying methods and having different refractive indices. The horizontal coordinate is given in the $x$-direction in Fig. \ref{fig:fig1}a. The aerogels were tagged with the mean refractive index ($n_{tag}$) measured at four corners of the aerogel tiles at $d$ = 5 mm. Because X-ray absorption measurements (Section IV) were performed along the positions from corner $B$ to corner $D$, the refractive index was measured at corner $B$ of each aerogel in detail; $x$ = 0 mm corresponds to the vertex $B$. The error bars correspond to the spread of the laser spots on the screen after they pass through the aerogels. The conventionally produced aerogels exhibited a flat refractive index distribution irrespective of the tagged refractive index. In contrast, the aerogels produced by pin-drying exhibited an inclined refractive index distribution; i.e., the refractive index increased by an amount ranging from 0.005 to 0.007 within the position range of the measurement as the position approached the corner of the aerogel tiles.

\section{X-ray Absorption Technique}
Second, we measured the uniformity of the aerogel density by an X-ray absorption technique. The evaluation consists of three steps:
\begin{itemize}
\item Elemental ratio analysis [X-ray fluorescence (XRF) analysis]
\item X-ray absorption measurement
\item Thickness measurement
\end{itemize}
The monoenergetic X-ray absorption by materials is described by the exponential attenuation law:
\begin{equation}
\label{eq:eq2}
I/I_0=\exp(-\mu _mx),
\end{equation}
where $I_0$ is the incident X-ray intensity, $I$ is the transmitted X-ray intensity, $\mu _m$ is the X-ray mass absorption coefficient, and $x$ is the mass thickness of the material. The mass thickness is defined as
\begin{equation}
\label{eq:eq3}
x=\rho t,
\end{equation}
where $t$ is the thickness of the material. The photon mass absorption coefficients for elements are given as functions of the photon energy in \cite{cite15}. In this study, the photon energy for X-ray absorption measurements was 8.04 keV. The mass absorption coefficient of compounds can be obtained to be a liner combination of each relevant element as follows:
\[ \mu _m=\sum_{i}w_i(\mu _m)_{i}, \]
where $w_i$ is the fraction by weight of the $i$th atomic constituent. The density of the materials is described using (\ref{eq:eq2}) and (\ref{eq:eq3}) as follows:
\[ \rho =\frac{1}{\mu _mt}\ln{\frac{I_0}{I}}. \]

\subsection{X-ray Fluorescence (XRF) Analysis}
To determine the aerogels' X-ray mass absorption coefficient, we performed XRF analysis. The coefficient is used to calculate the absolute density of aerogels. Instead of the aerogels listed in Table \ref{table:table1}, four similar aerogels prepared using DMF as a solvent were analyzed with an XRF instrument (ZSX100e, Rigaku). The X-ray mass absorption coefficient was obtained to be 33.9 $\pm $ 0.3 cm$^2$/g, which was an average from those aerogel samples. This corresponds to fractions by weights of 45.2\%, 48.7\%, and 6.0\% for silicon, oxygen, and carbon, respectively. The value of the coefficient was constant with changes in the refractive index of the aerogels. Our hydrophobic aerogels contain trimethylsiloxy groups [--OSi(CH$_3$)$_3$] added to the silica particles. In the XRF analysis, we selectively detected the weight content of carbon in addition to that of silica (silicon and oxygen). The instrument cannot detect hydrogen, but the effect of the element is negligible due to a small value of the mass absorption coefficient. The fractions by weights of the atomic elements were calibrated on the basis of the data obtained by analyzing quartz.

\subsection{X-ray Absorption Measurement}
X-ray absorption measurements were conducted on a NANO-Viewer (Rigaku) system, which consists of a monochromatic X-ray generator (Cu K$\alpha $, $\lambda $ = 1.54 \AA), a focusing multilayer optic with three slits, and a scintillation detector using a photomultiplier tube (R6249, Hamamatsu Photonics). We prepared an aerogel holder with a movable $X$--$Y$ stage. The $X$--$Y$ coordinate is given in Fig. \ref{fig:fig1}b. The holder has two one-sided slits on the $X$ and $Y$ rims to align the X-ray path with the measurement position on the aerogel tile upto an accuracy of 0.5 mm. By using the slits, we confirmed that the X-rays were focused on the holder within a diameter of approximately 1 mm. Except for the space where an aerogel was installed, the X-ray path was in vacuum ($<$100 Pa) to avoid X-ray scattering by air. The number of photons detected with the scintillation detector was counted with a scaler.

The scaler counts, after the X-rays pass through the aerogel, depend on the aerogel density. Depending on the density, the counts were accumulated for 5, 10, 20, or 40 s. The measurement was repeated five times at the same incident position, and the mean count was calculated as the transmitted X-ray intensity. We considered the standard deviation of the X-ray counts as the measurement error. Then, we measured the transmitted X-ray intensity at several positions along a diagonal line (between the corners $B$ and $D$) in the aerogels as shown in Fig. \ref{fig:fig1}b. To evaluate the incident X-ray intensity, counts without the aerogel were measured before and after the measurement with the aerogel. From the above data, the mass thickness [$x=(1/\mu _m )\cdot \ln(I_0/I)$] of the aerogels can be calculated as a function of the diagonal X-ray position $d$ (not shown). The uniformity in aerogel density can be estimated from the results of X-ray absorption measurements by combining with precise thickness measurements, which is shown in the next section.

\begin{figure}[t]
\centering
\includegraphics[width=3.5in]{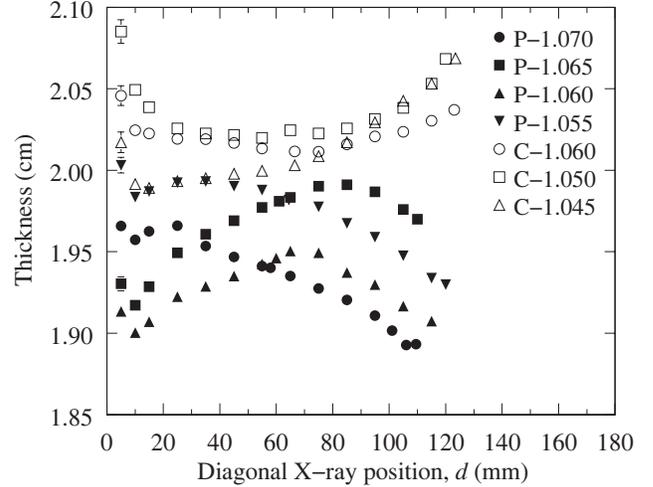}
\caption{Thickness as a function of the diagonal X-ray position $d$ for aerogels produced by pin-drying (filled symbols) or conventional (open symbols) methods.}
\label{fig:fig5}
\end{figure}

\subsection{Thickness Measurement}
The measurement of thickness is important to evaluate the uniformity in the aerogel density because the X-ray absorption is sensitive to changes in its thickness. Our study revealed that the thickness of the aerogels changes significantly, especially at the edge, owing to their meniscus geometry. To measure the exact thickness, we cut the hydrophobic aerogels along the X-ray scan lines with a water jet cutter. The thickness was measured with a universal measuring microscope (UMM200, Tsugami) consisting of a fixed microscope and a movable $x$--$y$ stage with a high-resolution (1 $\mu $m) digital position reversible counter (Nikon). By using the measuring microscope, we can observe the cross-sectional shape of the cut aerogels. It is crucial to align the position at which thickness is measured with that at which the X-ray intensity is measured. Considering the focused X-ray size (1 mm) and the positioning accuracy of the aerogels in the X-ray absorption and thickness measurements, we measured the thickness at $x$ = $d-0.5$, $d$, and $d + 0.5$ mm, where the center of X-ray beam was injected at $x = d$ (Fig. \ref{fig:fig1}b). The standard deviation of the thickness measured at these three positions was regarded as the measurement error of the thickness at $x = d$.

Fig. \ref{fig:fig5} shows thickness as a function of the diagonal X-ray position $d$ for the seven aerogels produced by the conventional or pin-drying methods. As shown in Fig. \ref{fig:fig1}b, the $x$-axis of Fig. \ref{fig:fig5} corresponds to the diagonal $BD$ (from 116 to 134 mm depending on the aerogel size). The thickness distribution of the conventionally manufactured aerogels reflected the natural meniscus geometry. In contrast, the thickness of the aerogels produced by pin-drying was convex upward. This is because further shrinking occurred at the edge of the aerogel tiles.

\subsection{Density Estimation}
\begin{figure}[t]
\centering
\includegraphics[width=3.5in]{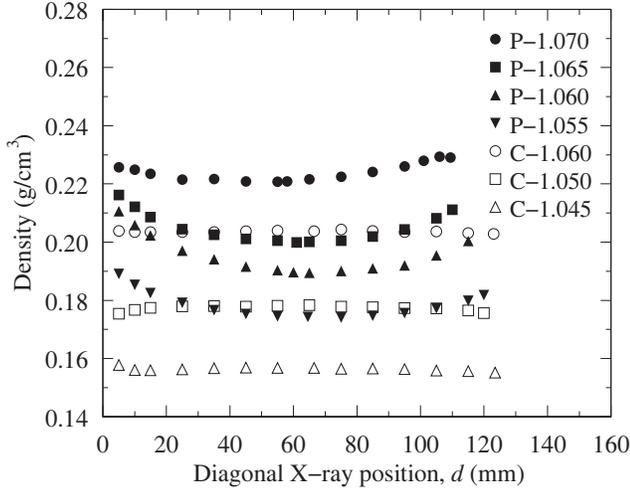}
\caption{Density as a function of the diagonal X-ray position $d$ for aerogels produced by pin-drying (filled symbols) or conventional (open symbols) methods.}
\label{fig:fig6}
\end{figure}

The density can be obtained by dividing the mass thickness by the thickness of the aerogels. Fig. \ref{fig:fig6} shows the density as a function of the diagonal X-ray position $d$ for the seven aerogels produced by the conventional or pin-drying methods. The measurement errors (standard deviations) are not shown because they are very small (maximum 0.002 g/cm$^3$). The conventionally manufactured aerogels had a flat density distribution. The density fluctuation was within 1\%. In contrast, the aerogels produced by pin-drying showed an inclined density distribution; i.e., the density increased by an amount ranging from 0.009 to 0.021 g/cm$^3$ as the position approached the corner of the aerogel tiles. For the aerogel P-1.060, the density increased by 11\% over that at the center of the aerogel tile. In addition, the density showed an asymmetric distribution. Shrinking of wet-gels clearly has negative effects on the uniformity in the aerogel density.

\begin{figure}[t]
\centering
\includegraphics[width=3.5in]{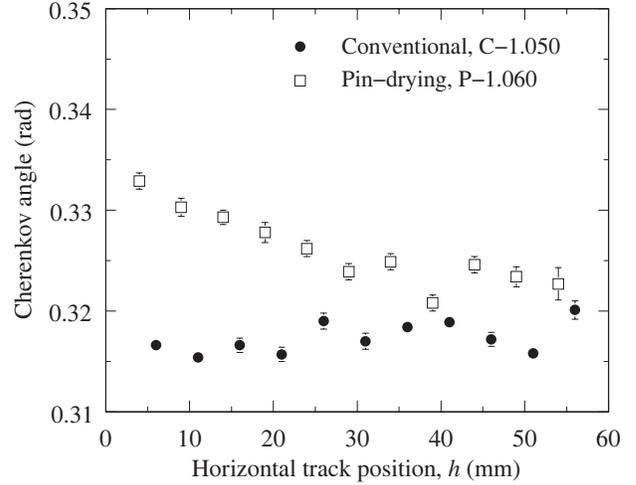}
\caption{Cherenkov angle as a function of the horizontal track position $h$ for the aerogels produced by the conventional (circles) or pin-drying (squares) methods. The horizontal axis corresponds to the $x'$-axis in Fig. \ref{fig:fig1}c.}
\label{fig:fig7}
\end{figure}

\section{Beam Test}
Finally, the aerogels' performance was confirmed by a beam test. At the Fuji test beam line at KEK, where 2 GeV/$c$ electron beam was available, we conducted a beam test to investigate the performance of the pin-dried aerogels and a prototype aerogel-RICH counter \cite{cite8} for the Belle II experiment. A 2 $\times $ 3 array of 144-channel position-sensitive hybrid avalanche photodetectors (HAPD) with a pixel size of 4.9 mm $\times $ 4.9 mm \cite{cite16}, which were read out by the electronics described in \cite{cite17}, was arranged in a light-shielded box. The distance between the upstream surface of an aerogel tile, used as a radiator, and the photodetection plane was set to 20 cm (i.e., the setup was that of a proximity focusing RICH counter). Electrons were tracked by two multiwire proportional chambers (MWPC) at the upstream and downstream ends of the light-shielded box; i.e., the beam incident positions on the aerogel tile can be measured by the MWPCs. We reconstructed the Cherenkov angle ($\theta _{ch}$) by analyzing each HAPD hit. The distribution of the local refractive index of the aerogel can be evaluated by measuring the Cherenkov angle at every beam incident position, because $\theta _{ch}$ is determined by the refractive index of the aerogel and the velocity $\beta $ of the electron:
\begin{equation}
\label{eq:eq4}
\cos\theta _{ch} = 1/(n\beta ),
\end{equation}
where $\beta $ = 1 for 2 GeV/$c$ electron.

To compare the aerogels produced by the conventional and pin-drying methods, the aerogels C-1.050 and P-1.060 were measured in the beam test. To facilitate the measurement, as shown in Fig. \ref{fig:fig1}c, we focused on the horizontal line connecting the midpoint of side $AB$ to the center of the aerogels ($x'$-axis). When the beam tracks passed across the horizontal line, the track events were divided into discontinuous eleven bins with 5-mm increments along the horizontal line, each of which has a 3 mm $\times $ 3 mm area. For each bin, the events were accumulated to be estimate the Cherenkov angle. The horizontal position $h$ is defined as the distance from the midpoint ($O'$) of side $AB$ to the center of the accumulated tracks  (Fig. \ref{fig:fig1}c). We used hits from one particular HAPD out of six since this HAPD covered bottom center part of the Cherenkov ring and always detected the ring efficiently even when we changed the injection point of electron beam. This method also reduced systematic effect coming from the HAPD difference. Fig. \ref{fig:fig7} shows the Cherenkov angle as a function of the horizontal track position $h$ for the two aerogels. We regarded the standard deviation of the mean ($\sigma_\theta /\sqrt{N}$) as the measurement error, where $\sigma_\theta $ is the standard deviation of the single photon Cherenkov angle distribution and $N$ is the number of Cherenkov photons accumulated over the experimental run ($N$ ranging from 50 to 400 depending on the horizontal track position). The conventionally manufactured aerogel exhibited an almost flat Cherenkov angle distribution. In contrast, the aerogel produced by pin-drying showed an inclined Cherenkov angle distribution; i.e., the Cherenkov angle increased by 0.008 rad as the position approached toward the side of the aerogel tile. This increase corresponds to an increase in the refractive index by 0.003.

\section{Discussion}
\begin{figure}[t]
\centering
\includegraphics[width=3.5in]{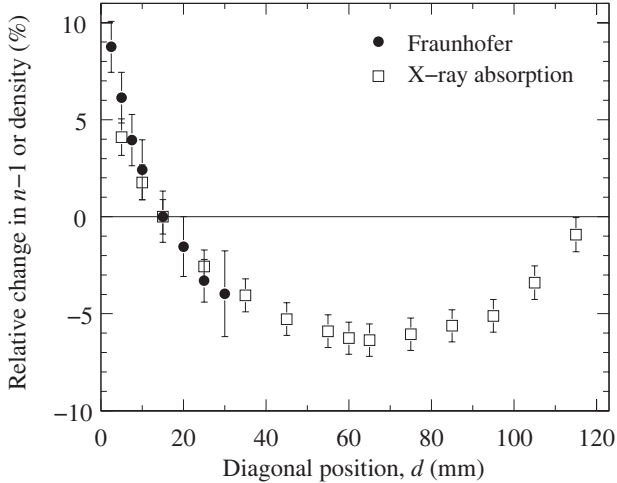}
\caption{Comparison of the measurement results of the Fraunhofer method (refractive index) and X-ray absorption technique (density). Relative change in $n-1$ (circles) and density (squares) of the aerogel P-1.060 are shown as functions of the diagonal position. The values of $n-1$ and density at $x$ = 15 mm were used as the standards for calculating the respective relative values.}
\label{fig:fig8}
\end{figure}

In this section, we discuss the consistencies in each measurement technique used to investigate the tile uniformity of aerogel P-1.060, produced by pin-drying. Fig. \ref{fig:fig8} shows the relative change in $n-1$ and density of the aerogel as a function of the diagonal position to compare the results of the Fraunhofer method with the X-ray absorption measurement. The refractive index and density at $x$ = 15 mm were used as standards to calculate the respective relative values. The results of the two measurements are consistent with each other. Although X-ray absorption measurement is a sensitive method for evaluating tile uniformity, it is a destructive technique to measure thickness accurately. In contrast, the Fraunhofer method is an easy tool for evaluating the uniformity around the corners of an aerogel tile. The comparison of these measurements reveals that the Fraunhofer method allows us to estimate the uniformity over a tile.

\begin{figure}[t]
\centering
\includegraphics[width=3.5in]{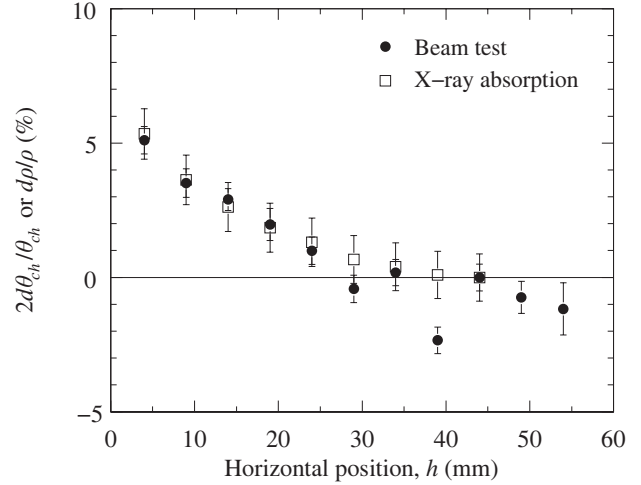}
\caption{Comparison of the measurement results of the beam test (Cherenkov angle) and X-ray absorption technique (density). Twice the value of the relative change in Cherenkov angle $2d\theta _{ch}/\theta _{ch}$ (circles) and the relative change in density $d\rho/\rho $ (squares) of the aerogel P-1.060 are shown as functions of the horizontal position. The Cherenkov angle and density at $x'$ = 44 mm were used as standards for calculating the respective relative values.}
\label{fig:fig9}
\end{figure}

For small $n-1$ (Cherenkov angle and $k\rho $), one can show that the relative change in Cherenkov angle is half of the relative change in density using (\ref{eq:eq1}) and (\ref{eq:eq4}): $2d\theta _{ch}/\theta _{ch}$ = $d\rho/\rho $. Fig. \ref{fig:fig9} shows twice the value of the relative change in Cherenkov angle $2d\theta _{ch}/\theta _{ch}$ and the relative change in density $d\rho/\rho $ of the aerogel as a function of the horizontal position to compare the result of the beam test with the X-ray absorption measurement. The Cherenkov angle and density at $x'$ = 44 mm (center of the aerogel tile) were used as standards for re-scaling each measurement value. The horizontal axis does not represent the diagonal but the horizontal position of the aerogel. Also the results of the two measurements are consistent with each other. The relative changes around the edge ($x'$ = 0 mm) of the aerogel tile were smaller than those around the corner ($x$ = 0 mm). 

However, as shown in Fig. \ref{fig:fig8}, the density difference between the center and the corner of the pin-dried aerogel was greater than 10\%, which is very large for our RICH counter. To solve this problem, we can use the highly transparent aerogels by removing their corners and edges with water jet cutters. Moreover, the pin-drying process must be improved to reduce the increase in density at the corners of pin-dried aerogels.

\section{Conclusion}
We can produce hydrophobic silica aerogels possessing excellent transparency using a novel pin-drying method. To characterize the uniformity of refractive index (i.e., density) in an individual monolith of the aerogel with $n$ ranging from 1.05 to 1.07, three methods were used: the Fraunhofer method (for refractive index measurement), an X-ray absorption technique (for density measurement), and Cherenkov imaging by a beam test using a RICH counter (for Cherenkov angle measurement). We confirmed that the aerogels produced by the conventional method were sufficiently uniform. However, we found a significant increase in density at the corners of the aerogels produced by pin-drying. To utilize the highly transparent pin-dried aerogels in our RICH counter, further ongoing studies for improving the tile uniformity affected by the pin-drying process are essential and will be reported elsewhere.

\section*{Acknowledgment}
M.T. and I.A. are grateful to the members of the Belle II A-RICH group for their assistance in the beam test. This work was partially supported by the Applied Research Laboratory and the Mechanical Engineering Center at KEK. This publication was supported in part by the Imaging Science Program of the National Institute of Natural Sciences (NINS) of Japan and the Space Plasma Laboratory at ISAS, JAXA.

\ifCLASSOPTIONcaptionsoff
  \newpage
\fi

\end{document}